\documentclass[argument]{aastex631}

\usepackage{hyperref}
\usepackage{amsmath}
\usepackage{enumerate}
\usepackage[utf8]{inputenc}
\usepackage{cleveref}
\usepackage{amssymb}
\usepackage[T1]{fontenc}
\usepackage{enumitem}
\usepackage{url}
\usepackage[misc]{ifsym}
\usepackage{natbib}
\received{June 11, 2022}
\revised{August 19, 2022}
\accepted{September 07, 2022}

\submitjournal{ApJ}

\graphicspath{{./}{figures/}}
\begin{document}

\title{Drift Rates of Narrowband Signals in Long-term SETI Observations for Exoplanets}
\shortauthors{Li et al.}

\author[0000-0002-1190-473X]{Jian-Kang Li}
\affiliation{Department of Astronomy, Yunnan University, Kunming 650500, China}
\affiliation{Department of Astronomy, Key Laboratory of Astroparticle Physics of Yunnan Province, Yunnan University, Kunming 650091, China}
\affiliation{Department of Astronomy, Beijing Normal University, Beijing 100875, China}
\author[0000-0002-5485-1877]{Hai-Chen Zhao}
\affiliation{Department of Astronomy, Beijing Normal University, Beijing 100875, China}
\affiliation{Institute for Frontiers in Astronomy and Astrophysics, Beijing Normal University, Beijing 102206, China}
\author[0000-0002-4683-5500]{Zhen-Zhao Tao}
\affiliation{Department of Astronomy, Beijing Normal University, Beijing 100875, China}
\affiliation{Institute for Frontiers in Astronomy and Astrophysics, Beijing Normal University, Beijing 102206, China}
\author[0000-0002-3363-9965]{Tong-Jie Zhang\href{mailto:tjzhang@bnu.edu.cn}{\textrm{\Letter}}}
\affiliation{Department of Astronomy, Beijing Normal University, Beijing 100875, China}
\affiliation{Institute for Frontiers in Astronomy and Astrophysics, Beijing Normal University, Beijing 102206, China}
\email{tjzhang@bnu.edu.cn}
\author[0000-0002-3464-5128]{Sun Xiao-Hui}
\affiliation{Department of Astronomy, Yunnan University, Kunming 650500, China}
\affiliation{Department of Astronomy, Key Laboratory of Astroparticle Physics of Yunnan Province, Yunnan University, Kunming 650091, China}
%% Note that the \and command from previous versions of AASTeX is now
%% depreciated in this version as it is no longer necessary. AASTeX 
%% automatically takes care of all commas and "and"s between authors names.

%% AASTeX 6.31 has the new \collaboration and \nocollaboration commands to
%% provide the collaboration status of a group of authors. These commands 
%% can be used either before or after the list of corresponding authors. The
%% argument for \collaboration is the collaboration identifier. Authors are
%% encouraged to surround collaboration identifiers with ()s. The 
%% \nocollaboration command takes no argument and exists to indicate that
%% the nearby authors are not part of surrounding collaborations.

%% Mark off the abstract in the ``abstract'' environment. 
\begin{abstract}
The Doppler shift of a radio signal is caused by the relative motion between the transmitter and receiver. The change in frequency of the signal over time is called drift rate. In the studies of radio SETI (Search for Extraterrestrial Intelligence), extraterrestrial narrow-band signals are expected to appear ``chirped'' since both the exoplanet and the Earth are moving. Such planet rotation and orbital revolution around the central star can cause a non-zero drift rate. Other relative motions between the transmitter and receiver, such as the gravitational redshift and galactic potential, are negligible. In this paper, we mainly consider the common cases that the drift rate is contributed by the rotations and orbits of the Earth and exoplanet in celestial mechanics perspective, and briefly discuss other cases different from the Earth-exoplanet one. We can obtain the expected pseudosinusoidal drifting result with long-term observations, shorter orbital periods of exoplanets. Exoplanets with higher orbital eccentricities can cause asymmetric drifting. The expected result should be intermittent pseudosinusoidal curves in long-term observations. The characteristics of pseudosinusoidal curves, as another new criterion for extraterrestrial signals, can be applied to long-term SETI reobservations in future research.

\end{abstract}
\keywords{\href{http://astrothesaurus.org/uat/2127}{Search for extraterrestrial intelligence(2127)}; \href{http://astrothesaurus.org/uat/211}{Celestial mechanics(211)}}
%% Keywords should appear after the \end{abstract} command. 
%% The AAS Journals now uses Unified Astronomy Thesaurus concepts:
%% https://astrothesaurus.org
%% You will be asked to selected these concepts during the submission process
%% but this old "keyword" functionality is maintained in case authors want

%% From the front matter, we move on to the body of the paper.
%% Sections are demarcated by \section and \subsection, respectively.
%% Observe the use of the LaTeX \label
%% command after the \subsection to give a symbolic KEY to the
%% subsection for cross-referencing in a \ref command.
%% You can use LaTeX's \ref and \label commands to keep track of
%% cross-references to sections, equations, tables, and figures.
%% That way, if you change the order of any elements, LaTeX will
%% automatically renumber them.
%%
%% We recommend that authors also use the natbib \citep
%% and \citet commands to identify citations.  The citations are
%% tied to the reference list via symbolic KEYs. The KEY corresponds
%% to the KEY in the \bibitem in the reference list below. 

\section{Introduction}
\par Radio frequency waves are ideal electromagnetic waves for interstellar communication since the radio photons cost less energy and the attenuation caused by interstellar medium absorption is lower (\citealt{1959Natur.184..844C}).  The Doppler shift of electromagnetic wave is caused by the relative motion of the transmitter and receiver. In radio SETI researches, we usually apply a ``drift rate'' as one of the significant parameters to distinguish extraterrestrial signals from signals generated on the Earth in radio observations. For a transmitter emitting signals with constant rest frequency and located on an exoplanet, the frequency of the received signals should change over time, mainly due to the rotations and the orbits around the central stars of the Earth and exoplanet. Such a change over time can be quantified as a non-zero drift rate. 
\par Most of the previous radio SETI observations and research projects only applied algorithms or used computing power to search for narrow-band signals with appropriate drift rates, but with little or no physical consideration. For example, the observations for various targets in the Breakthrough Listen (BL) project at different frequency ranges (e.g. \citealt{2013ApJ...767...94S}; \citealt{2020AJ....159...86P}; \citealt{2020AJ....160...29S}; \citealt{2021AJ....161...55M}; \citealt{2021AJ....162...33G}; \citealt{2022AJ....163..104F}), defined the maximum values of drift rates, then applied the ``tree deDoppler'' algorithm, which was 
modified from \cite{1974A&AS...15..367T}, to search for signals with drift rates fitting the characteristics of the exoplanet systems within the defined range. Other projects like SETI@home (\citealt{Anderson2002SETIhomeAE}) searched for signals drifting within the ranges of all physically possible situations. 
\par Recently, \cite{2019ApJ...884...14S} studied drift rate with astrophysical considerations. They proposed 9 possible factors that contribute to drift rate, presented a theoretical discussion on the effect of each factor, and calculated maximum drift rates for some representative systems. Although the theoretical discussion was detailed, covering the main physical cases, there were still some other factors that had not been taken into consideration, especially the orbit contributions from planets. In this paper, we mainly discuss the celestial mechanical factors that contribute to the relative motion of the transmitter and receiver, and derive a more accurate expression of drift rate for common Earth-exoplanet cases.
\par In Section \ref{sec:CMFDR}, we derive the expressions of drift rate over time mainly caused by planets' rotational and orbital motion, then illustrate the drift rates of some exoplanet systems with potential habitable planets. In Section \ref{Different Situations}, we discuss some special cases different from common Earth-exoplanet cases. Section \ref{Discussion} is a discussion on the drift characteristics that maximize the distinguishability of ETI signals. Finally, a summary and conclusion are given in Section \ref{Conclusion}.

\section{Celestial Mechanics Factors of Drift Rate}\label{sec:CMFDR}
\subsection{Derivation of Drift Rate}
\par
\begin{figure}[htb]
    \centering
     \includegraphics[scale=0.48]{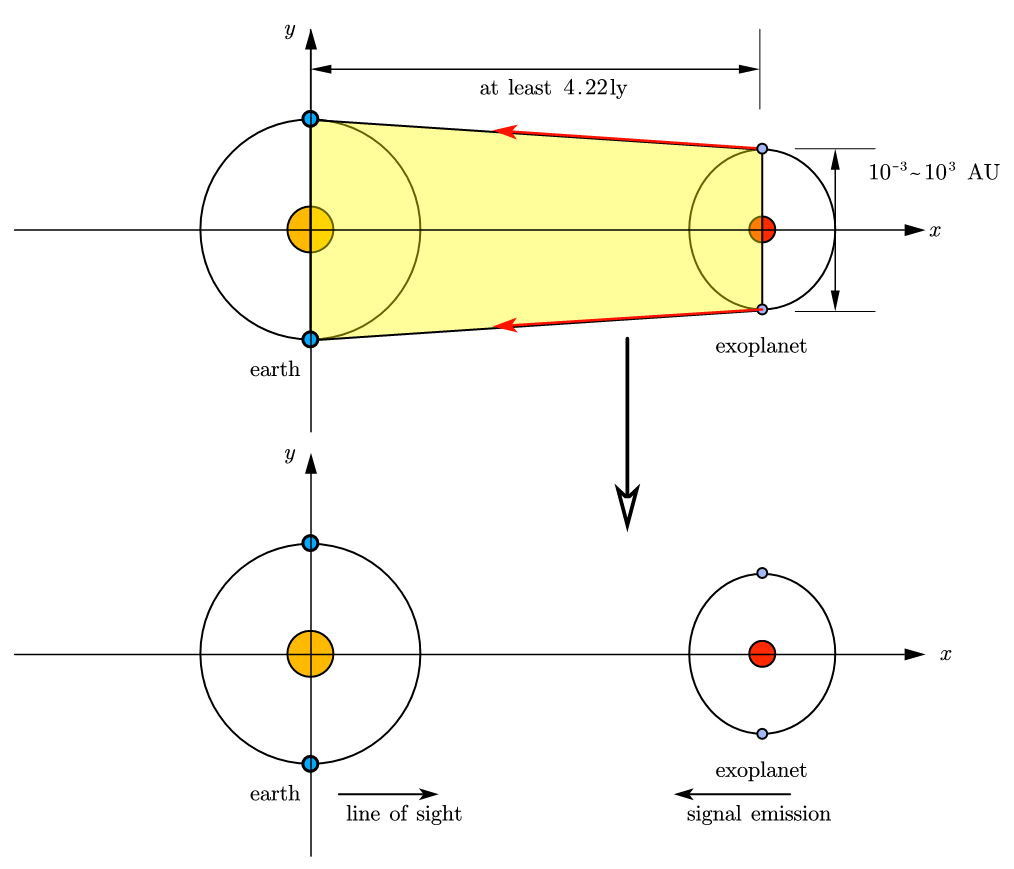}
     \caption{The coverage of emission from an exoplanet to the Earth (upper) and the simplified case for extremely small angle between the line of sight direction and the direction of emission (lower). The area filled with yellow is the location range where an observer on the Earth can detect signals from the exoplanet.}
     \label{lineinsight}
    \end{figure}

 The change in frequency and wavelength of light caused by the the relative motion of the signal source and observer can be explained by the Doppler effect, which can be expressed by redshift
\begin{equation}
z=\frac{\lambda-\lambda_0}{\lambda_0},
\end{equation}
where $z$ is the redshift, $\lambda$ represents the wavelength that observer receives and $\lambda_0$ represents the wavelength that the signal source emits. In general relativity, the Doppler effect should be written into relativistic form in Minkowski space: 
\begin{equation}
1+z=\frac{1+v_{s}\cos\alpha/c}{\sqrt{1-v^2_{s}/c^2}},
\label{relativisticwithangle}
\end{equation}
where $v_{s}$ is the relative velocity in the line-of-sight direction with respect to the signal source and observer, $c$ is the speed of light and $\alpha$ is the angle between the line-of-sight direction and the direction of emission in the observer's frame. The $\dfrac{1}{\sqrt{1-v^2_{s}/c^2}}$ term in equation (\ref{relativisticwithangle}) can be expended into Maclaurin series, then equation (\ref{relativisticwithangle}) can be changed into 
\begin{equation}
1+z=(1+v_{s}\cos\alpha/c)\sqrt{\sum_{n=0}^{\infty}{(v_{s}/c) ^{2n}}}.
\label{relativisticMac}
\end{equation}
In nonrelativistic limit (i.e. $v_{s}\ll c$), higher-order terms in Equation (\ref{relativisticMac}) can be omitted, and the approximation can be written as
\begin{equation}
z\approx \frac{v_{s}\cos\alpha}{c}.
\label{zapprox}
\end{equation}
The redshift in wavelength form can be changed into frequency form, then from Equation (\ref{zapprox}) we can write
\begin{equation}
\frac{1}{1+z}=\frac{\nu}{\nu_0}=\frac{c}{c+v_s\cos \alpha},
\label{frequencyform}
\end{equation}
where $\nu$ is the frequency that the observer receives and $\nu_0$ is the frequency that the signal source emits. We set the X-axis direction as the line-of-sight direction. In the case of an extraterrestrial transmitter and a receiver on the Earth, the orbital radius of an exoplanet is extremely small compared with the distance between the Earth and exoplanet (the closest planetary system to the Earth is Proxima Centauri system, which is about 4.22 lt-yr away, while the size of an orbit of exoplanet ranges from $10^{-3}$ to $10^3$ AU\footnote{The data of exoplanets can be seen in \url{https://exoplanetarchive.ipac.caltech.edu/}}), and we can only receive signals that cover the area of the Earth's orbit, which means that the parallax angle of the Earth's orbit can be regarded as zero, and consequently the line-of-sight direction can be regarded to be parallel to the direction of emission, but in opposite direction (i.e. $\alpha\approx180^{\circ}$) (See Figure \ref{lineinsight}). 
And for $v_s\ll c$, equation (\ref{frequencyform}) can be simplified by Maclaurin series omitting higher-order terms:
\begin{equation}
\nu=\nu_0+\nu_0\frac{v_{s}}{c}.
\label{freMac}
\end{equation} 
Then drift rate can be obtained by deriving equation (\ref{freMac}):
\begin{equation}
\dot{\nu}=\frac{v_{s}}{c} \frac{d\nu _0}{dt},
\label{driftrate}
\end{equation}
where $\dfrac{dv_s}{dt}$ is the total relative acceleration in the line of sight between the signal source and observer. For simplicity, we mainly discuss the situation in which the signal is a narrowband simple beacon that constantly emitted from the transmitter with a constant frequency in the reference frame of the receiver on Earth.
  \cite{2019ApJ...884...14S} wrote the relative radial acceleration into a sum of various terms in line-of-sight direction, $\sum\limits_i{\dfrac{dv_{s}}{dt}}$, and pointed out four main contributing terms: the rotation of the Earth, the orbit of the Earth, the rotation of the exoplanet where the transmitter is placed, and the orbit of the exoplanet. The contributions of the oblateness and topology of the Earth and exoplanet, the galactic potential, the gravitational redshift in the solar system are negligible. In addition, the following related effects can also be neglected in this work:
  \begin{enumerate}
  \item The rotational period of the Earth tends to be longer when approaching perihelion, while shorter when approaching aphelion (\citealt{1997mam..book.....M}). The same case can be applied to exoplanets with elliptical orbits.
  \item The tidal interaction with the Moon that causes gradual rotational deceleration in megayear timescale (\citealt{2000RvGeo..38...37W}). The same case can be applied to exoplanets if they have moons orbiting around them.
  \item The axial precession, astronomical nutation and polar motion of the Earth (\citealt{2010AAS...21547503S}). 
  \item The precession of the perihelion caused by the the curvature of spacetime(\citealt{1916AnP...354..769E}).
  \item The orbital motions around the Galactic Center (GC) of the solar system and extrasolar systems.
  \end{enumerate}
  \subsection{Rotation Terms of the Earth and Exoplanets}
  \par
\begin{figure}[htb]
    \centering
        \includegraphics[scale=0.45]{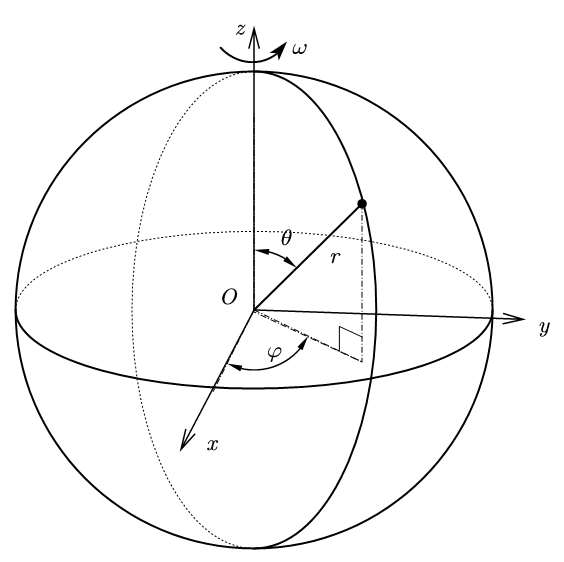}
        \caption{A certain point in spherical coordinate system, rotating at an angular velocity $\omega$}
        \label{spherecoordinate}
    \end{figure}

 For a receiver or transmitter on the surface of a rotating planet, applying a spherical coordinate system, the location of the receiver or transmitter can be given by radial distance, polar angle, and azimuthal angle written in $(r,\theta,\varphi)$ \footnote{ISO 80000-2:2019 is used in this paper. \url{https://www.iso.org/standard/64973.html}} (See Figure \ref{spherecoordinate}). 
  
\begin{figure}[h]
\centering
      \includegraphics[scale=0.45]{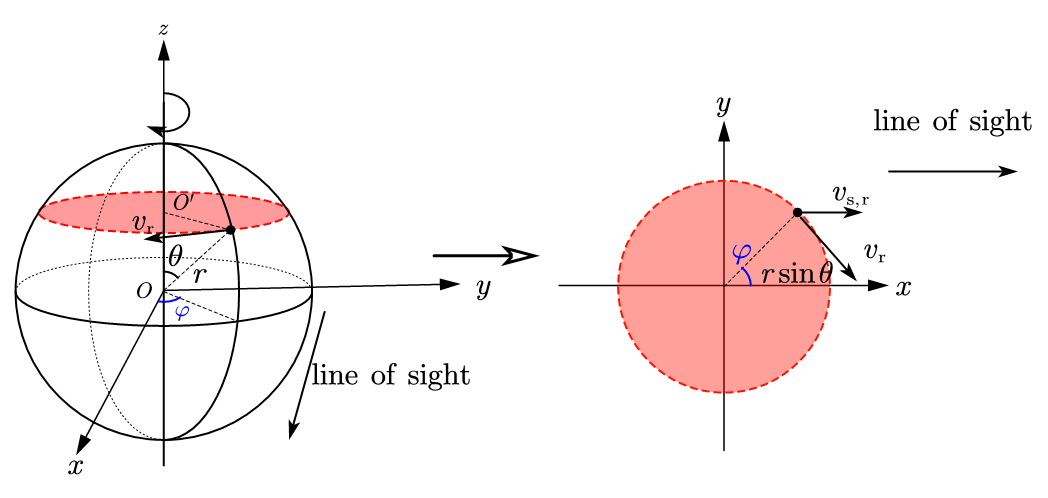}
      \caption{The motion of an object on a rotating planet and the simplified case.}
      \label{vecofrot}
  \end{figure}

  The motion of an object on a rotating planet can be simplified as uniform circular motion, since the oblateness and topology of the planet can be negligible, as well as the variation of the rotational period (See Figure \ref{vecofrot}), and then the velocity can be expressed by 
  \begin{equation}
 v_{s,r}=\frac{2\pi r\sin \theta}{P_{r}}\sin \varphi_r,
 \label{vecplanet}
  \end{equation} 
  where $P_{r}$ is the rotation period, $t$ represents time, and $\varphi=\frac{2\pi t}{P_{r}}$. Then, the acceleration in the line-of-sight direction can be derived by 
    \begin{equation}
    \frac{dv_{s,r}}{dt}=\frac{4\pi^2r\sin\theta}{P^2_{r}}\cos\varphi_{r}.
    \end{equation}

  \subsubsection{Rotation of the Earth}
  \par
\begin{figure}[htb]
       \centering
                \includegraphics[scale=0.5]{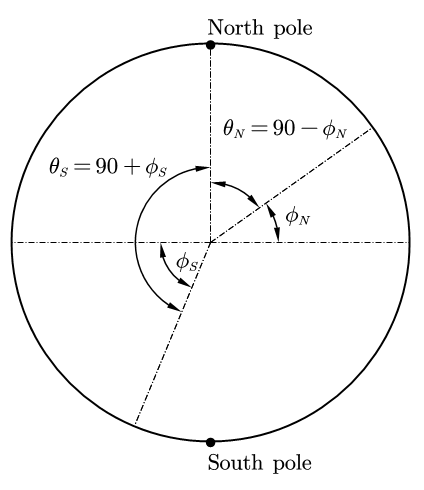}
                \caption{Conversion between latitude and polar angle}
                \label{latopolar}
      \end{figure}

 In most cases the receivers on the Earth are not located on the equator. The latitude $\phi$ can be converted to polar angle by $90^{\circ}\pm\phi$, plus for south semi-sphere and minus for north semi-sphere (See Figure \ref{latopolar}). For example, the latitude of 100m Robert C Byrd Green Bank Telescope (GBT): $38^{\circ}25'59''$ N can be converted to $\theta=51^{\circ}34'01''$, and the latitude of Parkes 64m Telescope (Parkes): $32^{\circ}59'52''$ S can be converted to $\theta=57^{\circ}00'08''$. Using $\oplus$ as the subscript of the Earth and replacing $r$ with $R$, for a receiver with polar angle $\theta_{re}$, the acceleration in the line-of-sight direction can be written as 
  \begin{equation}
  \frac{dv_{s,r,\oplus}}{dt}=\frac{4\pi ^2R_{\oplus}\sin \theta _{\text{re}}}{P_{r,\oplus}^{2}}\cos \varphi _{r,\oplus},
  \label{accEarth}
  \end{equation}
  where $R_{\oplus}$ is the radius of Earth. Equation (\ref{accEarth}) is the situation in which the direction in which transmitter points may not be parallel to the equator plane. For a target with decl. $\delta$, Equation (\ref{accEarth}) should be rewritten as
  \begin{equation}
  \frac{dv_{s,r,\oplus}}{dt}=\frac{4\pi ^2R_{\oplus}\sin \theta _{\text{re}}\cos\delta}{P_{r,\oplus}^{2}}\cos \varphi _{r,\oplus}.
    \label{accEarthdelta}
  \end{equation}
  \subsubsection{Rotation of Exoplanets}
  \par Similar to Equation (\ref{accEarthdelta}), the acceleration in line-of-sight direction for a transmitter on the exoplanet with polar angle $\theta_{tr}$ can be written as
  \begin{equation}
    \frac{dv_{s,r,p}}{dt}=\frac{4\pi ^2R_{p}\sin \theta _{\text{tr}}\cos \delta '_{\oplus}}{P_{r,p}^{2}}\cos \varphi _{r,p},
    \label{accexoplanet}
    \end{equation}
    where p is the subscript of exoplanet, $\theta_{tr}$ represents the polar angle of the transmitter, and $\delta'_{\oplus}$ is the decl. of Earth in the equatorial coordinate system of the exoplanet frame. To maximize the acceleration in the line-of-sight direction, the transmitter should be set on the equator of the exoplanet (i.e. $\theta_{\text{tr}}=90^{\circ}$), and the $\delta'_{\oplus}$ should be assumed to be 0. 
    \par The rotation period of an exoplanet is extremely tough to determine. \cite{2014Natur.509...63S} measured the rotation period of Beta Pictoris b, the first time for us to measure the rotation period of an exoplanet. \cite{2016ApJ...817..106B} used CO and H$_2$O spectra to measure the rotational velocity of HD 189733 b. However, the methods used in the above studies can be only applied to the cases of hot Jupiter, instead of Earth-like exoplanet. Another method to determine the rotation period of an exoplanet is tidal locking or spin–orbit resonance. Moreover, \cite{2017CeMDA.129..509B} implied that tidally locked exoplanets are common in Milky Way and habitable exoplanets are possible to be tidally locked by their host stars. TRAPPIST-1 system is a perfect example since all the seven planets are likely to have been tidally locked by TRAPPIST-1 (\citealt{2017Natur.542..456G}), and some of them are located in the habitable zone (\citealt{2017MNRAS.469L..26O}; \citealt{2020IJAsB..19..136A}). \cite{2012ApJ...761...83M} also pointed out that Gliese 581 d, a potentially habitable exoplanet, is extremely probable to be trapped into spin–orbit resonance.
  \subsection{Orbit Terms of the Earth and Exoplanets}
\begin{figure}[htb]
  \centering
       \includegraphics[scale=0.55]{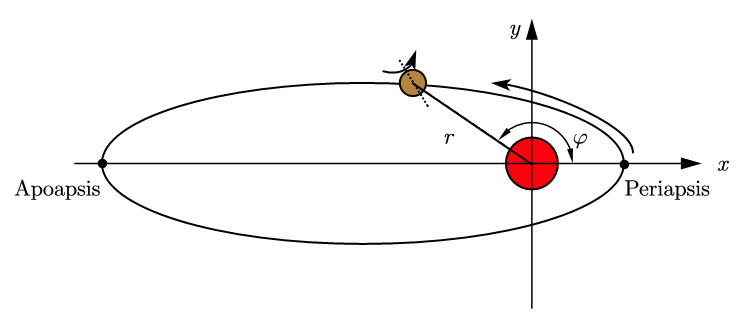}
       \caption{The orbit of a planet around its host star in polar coordinate system.}
       \label{polarcoordinate}
    \end{figure}

  The orbits of most planets are elliptical rather than perfectly circular. Applying a polar coordinate system, the orbit of a planet around its host star can be described by radial distance and polar angle written into ($r,\varphi$) (See Figure \ref{polarcoordinate}). Since the mass of a planet is several orders of magnitude smaller than its host star, the orbit of a planet can be considered as an ellipse around its host star.
    \par
\begin{figure}[htb]
         \centering
           \includegraphics[scale=0.55]{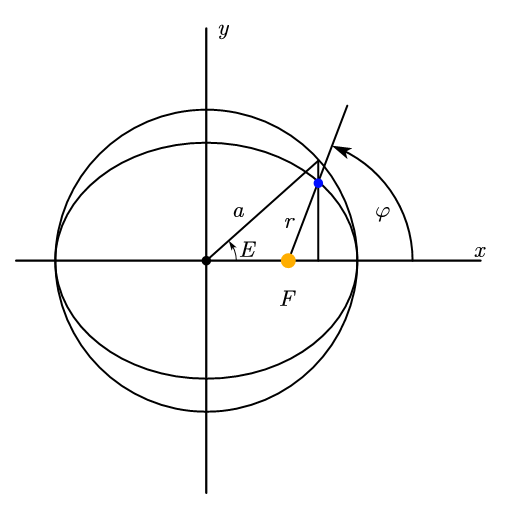}
            \caption{Eccentric anomaly is one of the angular parameters defined by a planet orbiting a central star along an elliptic Kepler orbit.}
                        \label{Eccentric anomaly}
        \end{figure}

 In a polar coordinate system, the radial distance can be written as 
\begin{equation}
r=\frac{a\left( 1-e^2 \right)}{1+e\cos \varphi_o},
        \label{radiusandpolar}
        \end{equation}
 where $e$ is the eccentricity of the orbit and $a$ is the length of the semimajor axis. The velocity vector and acceleration vector can be given by
      \begin{equation}
        \vec{v}=\dot{r}\hat{e}_r+r\dot{\varphi}_o\hat{e}_{\varphi},
        \end{equation}
         \begin{equation}
           \vec{a}=\left( \ddot{r}-r\dot{\varphi}_o^2 \right) \hat{e}_r+\left( r\ddot{\varphi}_o+2\dot{r}\dot{\varphi}_o \right) \hat{e}_{\varphi},
           \end{equation}     
 where $\hat{e}_r$ is the radial unit vector and $\hat{e}_{\varphi}$ is the angular unit vector. $v_{r}=\dot{r}$ and $a_r=\ddot{r}-r\dot{\varphi}^2$ are the radial velocity and radial acceleration, $v_{\varphi}=r\dot{\varphi}_o$ and $a_\varphi=r\ddot{\varphi}+2\dot{r}\dot{\varphi}$ are the circumferential velocity and circumferential acceleration, respectively. For the case of a central force field (gravitational field), only the radial acceleration exists, written as
 \begin{subequations}
 \begin{align}
a_r&=\ddot{r}-r\dot{\varphi}_o^2=\frac{GM_{\text{star}}}{r^2},
 \label{centralforce}\\
 a_{\varphi}&=r\ddot{\varphi}_o+2\dot{r}\dot{\varphi}_o=0,
 \end{align}
 \end{subequations}
 where $G$ is the gravitational constant and $M_{\text{star}}$ is the mass of central star. 
 Then, the velocity and acceleration in the line-of-sight direction can be written by 
    \begin{equation}
    v_{s,o}=\dot{r}\cos \varphi_o -r\dot{\varphi}_o\sin \varphi_o,
    \label{losvel}
    \end{equation}
    \begin{equation}
    \begin{aligned}
     \frac{dv_{s,o}}{dt}&=\left( \ddot{r}-r\dot{\varphi}_o^2 \right) \cos \varphi_o -(r\ddot{\varphi}_o +2\dot{r}\dot{\varphi}_o) \sin\varphi_\text{o}\\
     &=\frac{GM_{\text{star}}}{r^2}\cos\varphi_o.
    \end{aligned} 
    \label{losacc}
    \end{equation}
    Equation (\ref{radiusandpolar}) does not give the relation between polar coordinate and time, since the angular velocity of ellipses' orbital motion is not uniform. \cite{1609asno.book.....K} gave an alternative form, written as
    \begin{equation}
   \left\{ \begin{array}{l}
   	r=a\left( 1-e\cos E \right),\\
   	\varphi _o=2\arctan \left( \sqrt{\frac{1-e}{1+e}}\tan \frac{E}{2} \right),\\
   \end{array} \right. 
    \label{eccentric anomaly}
    \end{equation}
    where $E$ is the eccentric anomaly (See Figure \ref{Eccentric anomaly}). 
    The conversion between $E$ and $\varphi$ can be given by 
    \begin{equation}
    \cos \varphi_o =\frac{e-\cos E}{e\cos E-1}, \quad \sin E=\frac{\sqrt{1-e^2}\sin \varphi_o}{1+e\cos \varphi_o}.
    \end{equation}
    The Kepler equation (\citealt{1609asno.book.....K}) gives the relation between $E$ and $t$ by mean anomaly $M=2\pi t/P$, which can be written as
    \begin{equation}
    M=\frac{2\pi t}{P_o}=E-e\sin E.
    \label{mean}
    \end{equation}
    \cite{10.2307/2324547} gave a Bessel's solution of Kepler equation by extending Equation (\ref{mean}) into Fourier sine series:
    \begin{equation}
    E=\frac{2\pi t}{P_\text{o}}+\sum_{n=1}^{\infty}{\frac{2}{n}J_n\left( ne \right) \sin \left( \frac{2n\pi t}{P_\text{o}} \right)},
    \label{fourier}
    \end{equation}
    where $J_n\left( x \right) =\frac{1}{\pi}\int_0^{\infty}{\cos \left( nE-x\sin E \right) dE}$ is the Bessel integral in modern notation. Then, the radial velocity and circumferential velocity can be derived by Equations (\ref{eccentric anomaly}), (\ref{mean}) and (\ref{fourier}):
    \begin{footnotesize}
    \begin{subequations}
        \begin{gather}
        \dot{r}=\frac{2\pi a}{P_o}\left[ \sum_{n=1}^{\infty}{\frac{2}{n}J_n\left( ne \right) \sin \left( \frac{2n\pi t}{P_o} \right)} \right] \left[ 1+2\sum_{n=1}^{\infty}{J_n\left( ne \right) \cos \left( \frac{2n\pi t}{P_o} \right)} \right], \\
        r\dot{\varphi}_o=\frac{2\pi a\sqrt{1-e^2}}{P_o}\left[ 1+2\sum_{n=1}^{\infty}{J_n}\left( ne \right) \cos \left( \frac{2n\pi t}{P_o} \right) \right].
        \end{gather}
        \end{subequations}
    \end{footnotesize}
    
  \subsubsection{Orbit of the Earth}
  \par For a receiver located on Earth, the velocity is contributed by rotational velocity $v_{\text{r},\oplus}$ and orbital velocity $v_{\text{o},\oplus}$. However, the direction that points at the target is probably not parallel to the ecliptic plane. Therefore, the contribution from orbital velocity and should be its projection on equator. Rewriting Equations(\ref{losvel}), (\ref{losacc}) and (\ref{eccentric anomaly}) by using $\oplus$ as the subscript of Earth, the polar coordinate of Earth, velocity and acceleration in line-of-sight direction can be written as

  \begin{equation}
\left\{ \begin{array}{l}
	r_{\oplus}=\frac{a_{\oplus}\left( 1-e_{\oplus}^{2} \right)}{1+e_{\oplus}\cos \varphi _{o,\oplus}}=a_{\oplus}\left( 1-e_{\oplus}\cos E_{\oplus} \right),\\
	\varphi _{o,\oplus}=2\arctan \left( \sqrt{\frac{1-e_{\oplus}}{1+e_{\oplus}}}\tan \frac{E_{\oplus}}{2} \right),\\
\end{array} \right. 
        \label{radiusandpolarearth}
        \end{equation}
        \begin{equation}
        v_{s,o,\oplus}=(\dot{r}_{\oplus}\cos \varphi _{o,\oplus}-r_{\oplus}\dot{\varphi}_{o,\oplus}\sin \varphi _{o,\oplus})\cos\beta,
            \label{losvelearth}
        \end{equation}
  \begin{equation}
  \frac{dv_{s,o,\oplus}}{dt}=\frac{GM_{\odot}}{r^2_{\oplus}}\cos\varphi_{o,\oplus}\cos\beta.
   \label{orbitaccearth}
  \end{equation} 
  where $\beta$ is the ecliptic latitude of the target. The ecliptic coordinate of exoplanet is usually not provided; the NASA/IPAC Extragalactic Database provides a tool for the conversion of different astronomical coordinate systems \footnote{\url{https://ned.ipac.caltech.edu/coordinate_calculator}}. The input data are the right R.A. and decl., and the output is the ecliptic longitude and ecliptic latitude.
  \subsubsection{Orbit of Exoplanets}
  \par Since we have little or even no information about the axial tilt of exoplanets, here the angle between the rotational axis and orbital axis is set to be 0. For observers on the Earth, there may be an angle between the orbital plane and the plane perpendicular to the line of sight (i.e. the inclination of the exoplanet). Similar to Equations (\ref{radiusandpolarearth}), (\ref{losvelearth}) and (\ref{orbitaccearth}), the polar coordinate of the exoplanet, velocity and acceleration in the line-of-sight direction can be written as
 \begin{equation}
        \left\{ \begin{array}{l}
        	r_p=\frac{a_p\left( 1-e_{p}^{2} \right)}{1+e_p\cos \varphi _o,p}=a_p\left( 1-e_p\cos E_p \right),\\
        	\varphi _o,p=2\arctan \left( \sqrt{\frac{1-e_p}{1+e_p}}\tan \frac{E_p}{2} \right),\\
        \end{array} \right. 
         \label{radiusandpolarexoplanet}
         \end{equation}
         \begin{equation}
         v_{s,o,p}=(\dot{r}_{p}\cos \varphi _{o,p}-r_\text{p}\dot{\varphi}_{o,p}\sin \varphi _{o,p} )\sin i,
             \label{losvelexoplanet}
         \end{equation}
   \begin{equation}
   \frac{dv_{s,o,p}}{dt}=(\frac{GM_{star}}{r^2_\text{p}}\cos\varphi_{o,p})\sin i,
    \label{orbitaccexoplanet}
   \end{equation}
   where $M_{star}$ is the mass of the central star and $i$ is the inclination of the exoplanet. \cite{2014prpl.conf..715F} suggested the probability of a random orbital inclination between 30$^\circ$ and 90$^\circ$ is roughly 87\%.
  \subsection{Contributions of the Four Terms}
  \par Combining Equations (\ref{driftrate}), (\ref{accEarth}), (\ref{accexoplanet}), (\ref{orbitaccearth}) and (\ref{orbitaccexoplanet}) together, the drift rate caused by the rotations and orbits of the Earth and exoplanet can be expressed by
  \begin{equation}
  \dot{\nu}=\frac{\nu _0}{c} \left(\frac{dv_{\text{s,r,}\oplus}}{dt}+\frac{dv_{\text{s,r,p}}}{dt}+\frac{dv_{\text{s,o,}\oplus}}{dt}+\frac{dv_{\text{s,o,p}}}{dt}\right) , 
  \end{equation}
  where $\delta$ is the decl. of the exoplanet system. For simplicity, the initial $\varphi_r$ and $\varphi_o$ are assumed to start from zero. For an orbit with a relatively small eccentricity, the orbital motion can be approximated as uniform circular motion, written as
  \begin{equation}
  \frac{dv_{s,o}}{dt}=\frac{GM_\text{star}}{r^2}\cos\varphi_{o}\sin i,\quad \varphi_{o}=\frac{2\pi t}{P_o}.
  \label{fourterms}
  \end{equation}
It should be noticed that Equation (\ref{fourterms}) is applied to the signals observed by Earth that are within Milky Way. For extragalactic signals, other factors such as the redshift/blueshift and orbits around the galactic centers should be considered, making Equation (\ref{fourterms}) more complex.
  \subsection{Cases of Habitable Exoplanets}
  \par The Five-hundred-meter Aperture Spherical radio Telescope (FAST) is the largest single aperture radio telescope in the world (\citealt{2016RaSc...51.1060L}), covering 57\% of the celestial sphere. As the most sensitive radio telescope in the world (\citealt{2020RAA....20...64J};\citealt{2021RAA....21...18W}), FAST has carried out its first SETI observation by drift-scan survey (\citealt{2020ApJ...891..174Z}) and the first targeted SETI observation (\citealt{2022arXiv220802421T}). Since FAST has its unique advantages of large sky coverage and high sensitivity, we use FAST as an example of receivers on the Earth in the following calculations. Its latitude is $25^{\circ}39'10''$ N ($\theta=64^{\circ}20'50''$) and its operating frequency bands covers the range of 70 MHz – 3 GHz (\citealt{2016RaSc...51.1060L}). For transmitters on the equators of exoplanets, we take three known exoplanets: TRAPPIST-1 g, Kepler-438 b and Proxima Centauri b, which are potential habitable planets, as examples. The rotational and orbital parameters of Earth are listed in Table \ref{Parameters of Earth}, and the rotational and orbital parameters of three exoplanets are listed in Table \ref{Parameters of exoplanet}. The axial tilts of three exoplanets are assumed to be 0 for maximizing the drift rate.
    \begin{table}[htb]
      \centering
      \caption{The Rotational and Orbital Parameters of the Earth.}
      \begin{tabular}{cc}
      \hline
      \hline
      Parameters    & Earth                     \\ \hline
      $R$           & $6.3781\times10^6$m        \\ 
      $P_\text{r}$         & 86164.09890369732 s        \\ 
      $e$           & 0.0167086            \\ 
      $a$           & $1.4958023\times10^{11}$ m \\ 
      $P_\text{o}$         & $3.15581\times10^7$ s     \\ 
      $\varepsilon$ & $23.43641^{\circ}$       \\ 
      $M_{\oplus}$    & $1.989\times10^{30}$ kg    \\ \hline
      \end{tabular}
      \centering
      \\ \textbf{Note.} The parameters of the Earth and Sun are from \citet[radius]{2015arXiv151007674M}, \citet[rotational period]{1982A&A...105..359A}, \citet[eccentricity, semimajor axis and orbital period]{1994A&A...282..663S}, \citet[axial tilt]{1981SvA....25..263A}, \citet[mass of Sun]{2000asqu.book.....C} .
      \label{Parameters of Earth}
      \end{table}
  \subsubsection{TRAPPIST-1 g}
  \par TRAPPIST-1 g is one of the planets located inside the habitable zone of TRAPPIST-1 system (\citealt{2017MNRAS.469L..26O}). It was discovered by the Spitzer Space Telescope in 2017 (\citealt{2017Natur.542..456G}). Its lower eccentricity also indicates that the variation of its distance to the host star is small, giving a relatively stable climate to TRAPPIST-1 g. Since the size of TRAPPIST-1 planetary system is significantly small and all the planets are close enough to TRAPPIST-1, it is believed that all the planets are tidally locked by the host star (e.g. \citealt{2017Natur.542..456G}; \citealt{2018PNAS..115..260D}). Under such an assumption, the rotational period of TRAPPIST-1 g should be equal to its orbital period. For the transmitter operating on the surface of TRAPPIST-1 g at 1420MHz, the maximum drift rate we detected is 1.3294 Hz/s.
  \subsubsection{Kepler-438 b}
  \par Kepler-438 b was discovered by the Kepler Space Telescope in 2015. Although there are still debates on whether it is habitable because of the periodic radiation from its host star (\citealt{2016MNRAS.455.3110A}), it has still been studied and discussed in many researches of planetary habitability for its location inside the habitable zone of Kepler-438 and its high similarity to the Earth (\citealt{2017Ap&SS.362..146K}). Similar to most habitable planets orbiting around M dwarfs at close distances, Kepler-438 b is also believed to be tidally locked by its host star (\citealt{2017CeMDA.129..509B}). Under such an assumption, the rotational period of Kepler-438 b should be equal to its orbital period.  For the transmitter operating on the surface of Kepler-438 b at 1420 MHz, the maximum drift rate we detect is 0.709 Hz/s.
  \subsubsection{Porxima Centauri b}
   \par Proxima Centauri b is one of the closest exoplanets to our Solar system we have ever known, discovered by \cite{2016Natur.536..437A}. It is also located inside the habitable zone of Porxima Centauri. \cite{2016A&A...596A.111R} suggested that Porxima Centauri b is not likely to be tidally locked. Its rotation becomes stable in a 3:2 spin-orbit resonance for an eccentricity greater than 0.06–0.07, and a 2:1 spin-orbit resonance for the case that $e=0.16$. In this work, a 2:1 spin-orbit resonance is applied, for higher variation of the orbital radius. Since Proxima Centauri b was detected by Doppler spectroscopy, there is still no data for the inclination of Proxima Centauri b. In this work, the inclination is assumed as 90$^{\circ}$ for maximizing the drift rate. For the transmitter operating on the surface of Proxima Centauri b at 1420 MHz, the maximum drift rate we detect is 2.1536 Hz/s.
\begin{table}[htb]
\centering
\caption{The Rotational and Orbital Parameters of TRAPPIST-1 g, Kepler-438 b and Porxima Centauri b.}
\begin{tabular}{cccc}
\hline
\hline
Parameters & TRAPPIST-1 g                              & Kepler-438 b                        & Porxima Centauri b                \\ \hline
$R$        & $1.148_{-0.033}^{+0.032}R_{\oplus}$       & $1.12\pm0.16R_{\oplus}$             & $1.30_{-0.62}^{+1.20}R_{\oplus}$  \\
$P_\text{r}$\tablenotemark{a}       & $12.352446$ days                 & 35.23319 days                       & 5.5593 days                        \\
$e$        & $0.00208\pm0.00058$                       & $0.03_{-0.03}^{+0.01}$              & 0.16                              \\
$a$        & $0.04687692\pm 3.2\times10^{-7}\text{AU}$ & $0.166^{+0.051}_{-0.042}$AU         & $0.04856\pm0.00030$AU             \\
$P_\text{o}$      & $12.352446\pm 0.000054$ days            & $35.23319_{-0.00025}^{+0.00029}$ days                        & $11.1868_{-0.0031}^{+0.0029}$ days \\
$M_\text{star}$ & $0.0898\pm 0.0023M_{\odot}$               & $0.544^{+0.041}_{-0.061}M_{\odot}$  & $0.1221\pm0.0022M_{\odot}$        \\
$i$        & $89.721_{-0.026}^{+0.019}(^{\circ})$                 & $89.86^{+0.14} _{-0.32} (^{\circ})$ & $90^{\circ}$\tablenotemark{b}                      \\
$\alpha$   & $23^{h} 06^m 29.283^s$                    & $18^h 46^m 34.9970^s$               & $14^h 29^m 42.94853^s$            \\
$\delta$   & $-05^{\circ} 02' 28.59''$                 & $+41^{\circ} 57' 03.9233''$         & $-62^{\circ} 40' 46.1631''$       \\
$\beta$\tablenotemark{c}    & $0^{\circ} 38' 0.82''$                     & $64^{\circ} 33' 55.48''$             & $-44^{\circ} 45' 47.87''$      \\
\hline   
\end{tabular}
\\ \textbf{Note.} The parameters of TRAPPIST-1 and TRAPPIST-1 g are from \citet[radius, eccentricity and semimajor axis]{2018A&A...613A..68G}, \citet[orbital period and mass of TRAPPIST-1]{2021PSJ.....2....1A}, \citet[inclination]{2018MNRAS.475.3577D}, \citet[R.A. and decl. of the system]{2003yCat.2246....0C}. The parameters of Kepler-438 and Kepler-438 b are from \citet[radius, eccentricity, semimajor axis, orbital period, mass of Kepler-438 and inclination]{2015ApJ...800...99T}, \citet[R.A. and decl. of the system]{2018A&A...616A...1G}. The parameters of Porxima Centauri and Porxima Centauri b are from \citet[radius]{2020AJ....159...41T}, \citet[eccentricity]{2016A&A...596A.111R},  \citet[orbital period and semimajor axis]{2022A&A...658A.115F}, \citet[mass of Porxima Centauri]{2017A&A...598L...7K}, \citet[R.A. and decl. of the system]{2007A&A...474..653V}.
\tablenotetext{a}{The rotational periods of TRAPPIST-1 g and Kepler-438 b are considered as the same as their orbital periods under the assumption of tidal locking, and the rotational period of Porxima Centauri b should be half of its orbital period for spin-orbital resonance of 2:1 (\citealt{2016A&A...596A.111R}).}
\tablenotetext{b}{The inclination of Porxima Centauri b is assumed to be 90$^{\circ}$ for maximizing the drift rate.}
\tablenotetext{c}{The ecliptic latitudes of the exoplanet systems is converted by the Coordinate Calculator of NASA/IPAC Extragalactic Database.}
\label{Parameters of exoplanet}
\end{table}

\subsubsection{Calculation Results}
There is a special waveband extending from 1420 to 1720 MHz that embraces the hydrogen and hydroxyl spectral lines in SETI research that is called the ``water hole'' (\citealt{2009seex.book.....R}). Table \ref{calculation} lists the calculation results for the maximum drift rate created by Earth and the above three exoplanets at the frequency of 1420 MHz.
\begin{table}[htb]
\centering
\caption{Calculation Results for the Maximum Drift Rate Created by Earth and the Above Three Exoplanets at the Frequency of 1420 MHz. }
\begin{tabular}{cccc}
\hline
\hline
Exoplanet          & Rotation terms & Orbit terms & $\dot{\nu}$ \\ \hline
TRAPPIST-1 g       & 0.1454        & 1.1804     & 1.3258     \\ 
Kepler-438 b       & 0.1078        & 0.6012     & 0.7090      \\ 
Porxima Centauri b & 0.0732        & 2.0804     & 2.1536      \\ \hline
\end{tabular}
\\ \textbf{Note.} Rotation terms are the drift rate created by the rotation of Earth and exoplanets. Orbit terms are the drift rate created by the orbit of Earth and exoplanets.
\label{calculation}
\end{table}
\par
\begin{figure}[htb]
         \centering
                  \includegraphics[scale=0.6]{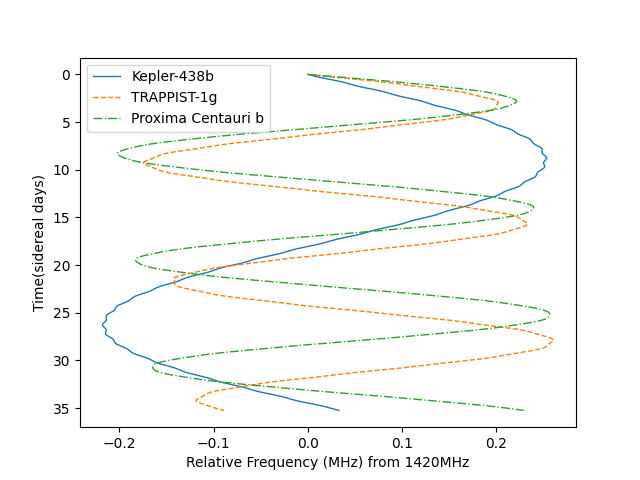}
                  \caption{Pseudosinusoidal relative frequency curves of long-term observations for TRAPPIST-1 g, Kepler-438 b and Proxima Centauri b.}
                  \label{exoplanets}
       \end{figure}

 Figure \ref{exoplanets} illustrates the frequency change from 1420MHz of the above discussed exoplanets. At the beginning of the simulated long-term observation, all terms of accelerations are at maximum while the velocity terms are at minimum. The slope of the curve is the corresponding drift rate at a certain point.
 \section{Some Different Cases}\label{Different Situations}
\subsection{Interferometers}
 \par The above discussions are easily applied to the observations of single dish. For small telescope arrays, Equation (\ref{fourterms}) can be also applied to their observations since the latitudes of each telescope are the same. For large telescope arrays or interferometers with long baseline, such as Square Kilometer Array (SKA) and Event Horizon Telescope (EHT), equation (\ref{fourterms}) can not be applied to the observations, since the locations of the telescopes (arrays) are significantly different. And the Earth's rotation term $\frac{dv_{s,r,\oplus}}{dt}$ should be correspondingly changed.
 \par
\begin{figure}[htb]
  \centering
          \centering
           \includegraphics[scale=0.5]{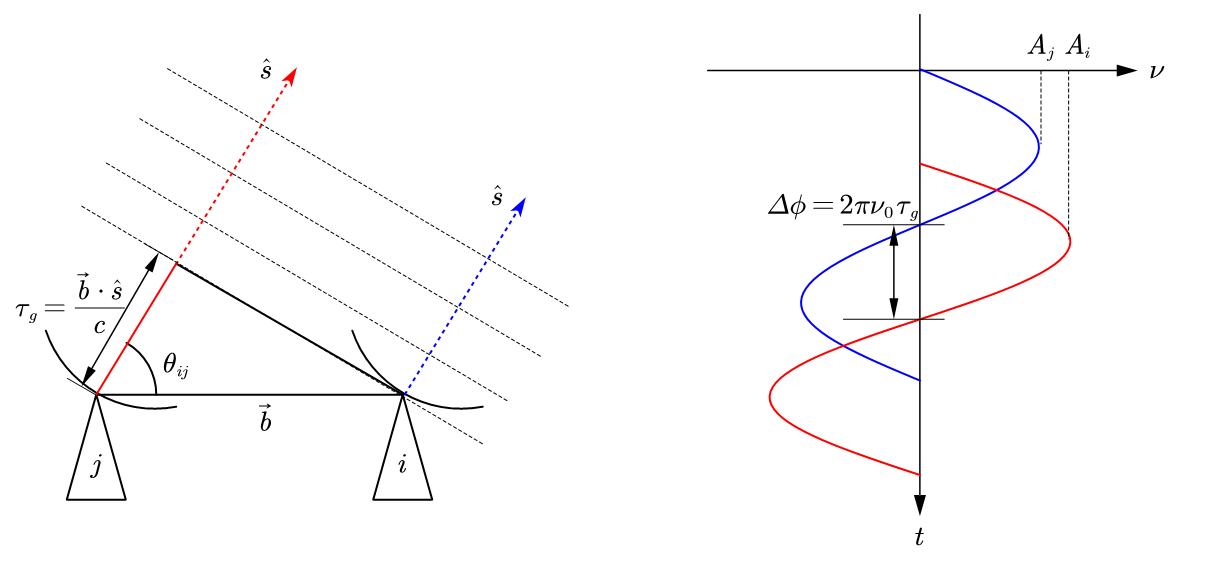}
           \caption{Left: extraterrestrial signal received by a pair of telescopes pointing at the same direction $\hat{s}$. The baseline vector $\boldsymbol{b}$ points from the $j$th telescope to $i$th telescope. Right: frequency drifts of the signal received by the two telescopes. The different locations of two telescopes cause the geometric delay of the signal.}
           \label{telescope}
\end{figure}

 For large telescope arrays or interferometers with long baseline, Equation (\ref{accEarthdelta}) should be rewritten as 
 \begin{equation}
  \frac{dv_{{s,r,}\oplus}}{dt}=\frac{4\pi^2R_{\oplus}\sin\theta_{\text{re},i}\cos\delta}{P_{{r},\oplus}^2}\cos\varphi_{{r},\oplus},
    \label{accEarthdeltaint}
  \end{equation}
  where $\theta_{\text{re},i}$ is the polar angle of the $i$th telescope.  If a signal is actually a extraterrestrial signal, when a pair of telescopes receive this signal, as Figure \ref{telescope} illustrates, the wave of the signal should travel an extra distance $\boldsymbol{b}\cdot \hat{s}=b\cos \theta _{ij}$, and the output frequency curves of the two telescopes should have a geometric delay
  \begin{equation}
      \tau _{g,ij}=\frac{b\cos \theta _{ij}}{c},
      \label{geometricdelay}
  \end{equation}
  where $\boldsymbol{b}$ is the baseline vector with length $b$, $\hat{s}$ is the  unit vector in the direction that the telescopes point at, and $\theta _{ij}$ is the angle between $\boldsymbol{b}$ and $\hat{s}$. Such geometric delay should cause a phase difference between two curves $\Delta\phi_{ij}$:
  \begin{equation}
      \Delta\phi _{ij}=2\pi\nu_0\tau_{g,ij}.
      \label{phasedif}
  \end{equation}
  The amplitudes of the curves and the maximum drift rate may have small difference between each other since the telescopes may be located at different latitudes (i.e. the polar angle $\theta_{\text{re},i}$ and $\theta_{\text{re},j}$ may not be equal). The criteria of phase and amplitude differences can be used to determine whether a signal is an ETI signal. 
 \subsection{Spacecrafts, Space Probes and Interstellar Probes}
 \par Transmitters that emit ETI signals may not be located on the surface of an exoplanet. A transmitter may be launched as a satellite orbiting an exoplanet or a host star and emits signals in all directions. For the case that a satellite orbits an exoplanet, its velocity and acceleration can be written as
 \begin{equation}
 v_{s,t}=\sqrt{\frac{GM_{p}}{\left( R+h \right)}}\sin \varphi _r,
 \end{equation} 
 \begin{equation}
 a_{s,t}=\frac{GM_{p}}{\left( R+h \right) ^2}\cos \varphi _r,
 \end{equation} 
 where $M_{p}$ is the mass of the exoplanet, $h$ is the height of the satellite. The orbital period can be calculated by the Kepler's third law:
 \begin{equation}
 \frac{\left( R+h \right) ^3}{P^2}=\frac{GM_{p}}{4\pi ^2}.
 \end{equation}
 For the case in which a satellite orbits a host star, the rotation term can be omitted. Similar to the orbit terms of the Earth and exoplanets, the nonuniform motion of an elliptic orbit should also be taken into consideration.
 \par A space probe launched into the space is able to escape its planetary system, such as Voyager 1. Interstellar probes have already escaped their planetary systems and are travelling in the interstellar space. For these two cases, both the rotation terms and orbit terms can be omitted. What we need to consider is the velocity in the line of sight. The velocity of an interstellar probe is significantly greater than space probes. \cite{2016JBIS...69...40L} proposed an envisioned interstellar probe with a laser phased array as energy propulsion. The maximum velocity can reach more than $c$/4 in a few minutes. And the acceleration should gradually decline from $\sim10^5$ m/s$^{2}$ to $\sim10^{-11}$ m/s$^2$ after reaching the maximum velocity. Such a decrease of acceleration can cause a remarkable drift rate, which gradually decreases to a negligible level over time. In comparison, Voyager 1 spacecraft was launched in 1977, its velocity has been increased to 17 km/s after 37 yr of flight.  For a spacecraft with a slow and uniform velocity like Voyager 1, its motion can cause a Doppler shift of an emitted signal, but with zero drift rate, since the acceleration is zero for uniform motion.    
\subsection{Signal from the Galactic Center}\label{GC}
\par The GC is considered to contain a large number of habitable planets owing to the higher number density of stars. The model proposed by \cite{2011AsBio..11..855G} and the following work by \cite{2015AsBio..15..683M} and \cite{2021AJ....162...33G} inferred that the GC is an ideal region for emergence of life and for technosignature detection. \cite{2019AJ....158..117C} suggested that higher densities of stars and habitable planets are beneficial to the development of advanced space-faring societies. Simulations of \cite{2022MNRAS.513...90D} also revealed that about half of planets orbiting around the Sagittarius A*  cluster (S stars), a cluster of stars in close orbit around Sagittarius A* (Sgr A*), can survive in the unstable environment in the GC.
\par Assuming that space-faring societies have formed in the GC. They would communicate and travel between planets orbiting S stars. The motion of S stars can be regarded as elliptical orbital motion. Until now, S4716 is the closest star orbiting Sgr A* that we found, with estimated semimajor axis lenght of $1.93\pm0.02\times10^{-3}$ pc, eccentricity of $0.756\pm0.02$ and the orbital period of 4 years (\citealt{2022ApJ...933...49P}). The velocity and acceleration of S4716 can be calculated by Equation (\ref{losvel}) and (\ref{losacc}), and the maximum acceleration of S4716 should be 2.5753 m/s$^2$. However, we have no information whether there are planets actually orbiting the S-stars and their orbital parameters. For simplicity, we assume that the orbital and rotational contributions of the exoplanets orbiting S-star can be neglected. For our solar system, the velocity orbiting GC is $\left( 233.6\pm 2.8 \right) \times 10^3$ m/s at a distance to GC $R_0=8.112\pm0.031$ kpc (\citealt{2019ApJ...870L..10M}), resulting in an orbital period of $P=2.13754\times 10^8$ years. And the acceleration toward GC is $\left( 2.32\pm 0.16 \right) \times 10^{-10}$ m/s$^{2}$ (\citealt{2021A&A...649A...9G}). Hence the contribution from the orbit of solar system can be neglected even in galactic scale. For a transmitter operating on the planet orbiting the S4716, if the the orbital and rotational contribution of the planet can be neglected, the the maximum drift rate caused by the orbit around Sgr A* of S4716 is 12.1898 Hz/s. Since the orbits of the stars of Sgr A* cluster are different from each other, the drift rates of the signals we receive should be different. High drift rate and the drift rate difference of the signal can be the criteria of ETI signals from the GC.
\par The complex environment of the GC also provide the opportunity to detect other natural astrophysical phenomena, such as pulsars, magnetars, radio transients or other mysterious phenomena. The pulsars in the GC can give us access to study the magnetoionic material, the gravitational potential, and star formation in the GC (\citealt{1997ApJ...475..557C}). The study of magnetars in the GC can provide us some key information about the origin of fast radio bursts. \cite{2021ApJ...920...45W} found a highly polarized radio source ASKAP J173608.2-321635 in the GC, and our current theories can not fully explain the observational results for this source. The BL observation of \cite{2021AJ....162...33G} also included these natural astrophysical phenomena as one of the science goals.

\subsection{Radio Frequency Interference}
\par A signal can be defined as narrowband signal if its bandwidth $B$ are significantly smaller than the central frequency $\nu_0$ ($B/\nu_0\ll1$), otherwise it should be defined as a broadband signal. Radio frequency interference (RFI) consists of artificial signals emitted from electronic devices. Narrowband RFI is the most common kind of RFI, which can be divided into zone RFI and drifting RFI. Zone RFI maintains stable frequency, mainly caused by televisions, radio broadcasts, cell phones and satellites, while the frequency of drifting RFI varies over time, mainly caused by mobile devices (\citep{2020ApJ...891..174Z}). Narrowband RFI can be expressed as the sum of sinusoids (\citealt{2013ITSP...61.4562S}), written into
\begin{equation}
I_{NB}\left( t,\tau \right) =\sum_{n=1}^N{a_n\left( t,\tau \right)}e^{i\left( 2\pi f_nt+\phi _n \right)},
\label{INB}
\end{equation}
where $a_n\left( t,\tau \right)$, $f_n$ and $\phi_n$ are the complex envelope, frequency, and initial phase of the n-th interference, $t$ and $\tau$ represent fast time and short time respectively. Broadband RFI is not frequently seen in observations, but it can contaminate the frequency bins and signals if it appears in the observational data. There are two types of broadband RFI: chirp-modulated (CM) broadband RFI and sinusoidal-modulated (SM) broadband RFI (\citealt{2019RemS...11.1654F}), CM broadband RFI can be expressed as:
\begin{equation}
I_\text{CM}\left( t,\tau \right) =\sum_{n=1}^N{a_n\left( t,\tau \right) e^{i\left( 2\pi f_nt+\pi \gamma _nt^2 \right)}},
\label{ICM}
\end{equation}
where $a_n\left( t,\tau \right)$, $f_n$, $\gamma_n$ indicate the complex envelope, frequency, and the chirp rate of the n-th interference, while SM broad-band RFI can be expressed as:
\begin{equation}
I_\text{SM}\left( t,\tau \right) =\sum_{n=1}^N{a_n\left( t,\tau \right) e^{i\beta_n\sin\left( 2\pi f_nt+\phi_n \right)}},
\label{ISM}
\end{equation}
where $a_n\left( t,\tau \right)$, $\beta_n$, $f_n$, $\phi_n$ indicate the complex envelope, modulation factor, frequency, and the initial phase of the n-th interference, respectively. 
\par
\begin{figure}[htb]
  \centering
          \centering
           \includegraphics[scale=0.6]{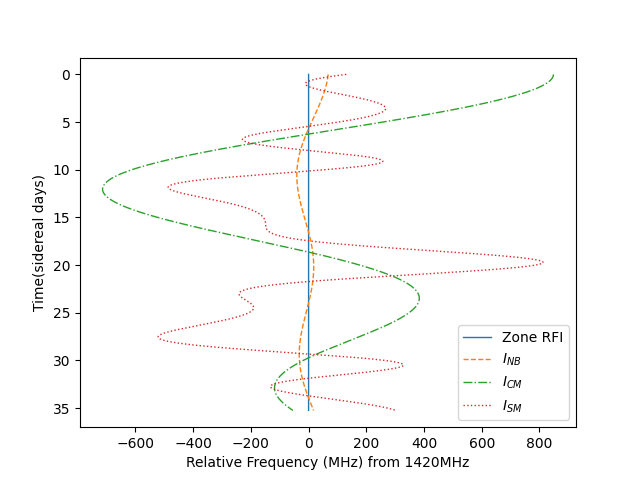}
           \caption{Summed pseudosinusoidal relative frequency curves of long-term observations for uninterrupted random-generated zone RFI, narrowband RFI, CM broadband RFI and SM broadband RFI.}
           \label{RFI}
\end{figure}

 Such sums of complex sinusoids should be similar to the results of extraterrestrial intelligence (ETI) signals, making it difficult for us to separate ETI signals and RFI. For RFI consisting of stationary $I_\text{st}$ and drifting part $I_\text{ns}$: $I(t)=I_\text{st}+I_\text{dr}(t)$, the received signal may appear a pseudosinusoidal curve if the drifting RFI follows Equation (\ref{INB}), Equation (\ref{ICM}), or Equation(\ref{ISM}). An exception is the zone RFI created by stationary sources on the Earth, since there is no radial acceleration causing drift, resulting in a zero drift rate (See Figure \ref{RFI}). The result in Figure \ref{RFI} is a hypothetical situation that RFI is uninterrupted for about 35 days. Actually, the duration of RFI is probably within microseconds to hours. 

\par Some observational or computational methods have been applied to RFI removal. A common observational method in SETI research is the ``On-Off" observational strategy, which was proposed in the BL project by \cite{2017PASP..129e4501I}, and has been applied to the following observations and researches in the project. Each observation for primary target is followed by an observation for off-source target which is different from the primary target. The sample of off-source target stars is from the \textit{Hipparcos} catalog. ETI signals are expected to only appear in ``On" observations while RFI should appear in both ``On" and ``Off" observations. Such observational strategy is often applied to single dish observations since it is relatively easy for single dish to switch observation targets, but difficult for interferometric arrays. Other specific methods such as cross correlation (\citealt{Abdelgelil2019RFICF}) and spatial filtering (\citealt{Veen2016SpatialFO}) are uesd by interferometric arrays use to mitigate RFI. Nebula is a back-end data analysis system originally developed and used in SETI@home \footnote{\url{https://setiathome.berkeley.edu/nebula/index.php}} for RFI removal. It was also applied to the first SETI observations of FAST (\citealt{2020ApJ...891..174Z}). The hits in different frequency bins is marked and removed after the full set of raw data goes through the identification methods of zone FRI, drifting RFI and multibeam RFI in the process of Nebula RFI removal. More than 99\% RFI are removed by Nebula, and the rest are removed by k-Nearest-Neighbor (KNN) algorithm. 
\section{Discussion}\label{Discussion}
\par The forms of received signals, if they are actually from ETI, should appear to be pseudosinusoidal curves after removing RFI from the data, due to the rotational and nonuniform orbital motions of the Earth and exoplanets. Therefore, we can distinguish ETI signals by the characteristics of pseudosinusoidal curves. Since drift rates are caused by the periodic motions of planets, in order to search for such signals in our observations, we need to focus on the observational periods and the periods of the exoplanets, especially the orbital ones.
\par
\begin{figure}[htb]
          \centering
           \includegraphics[scale=0.18]{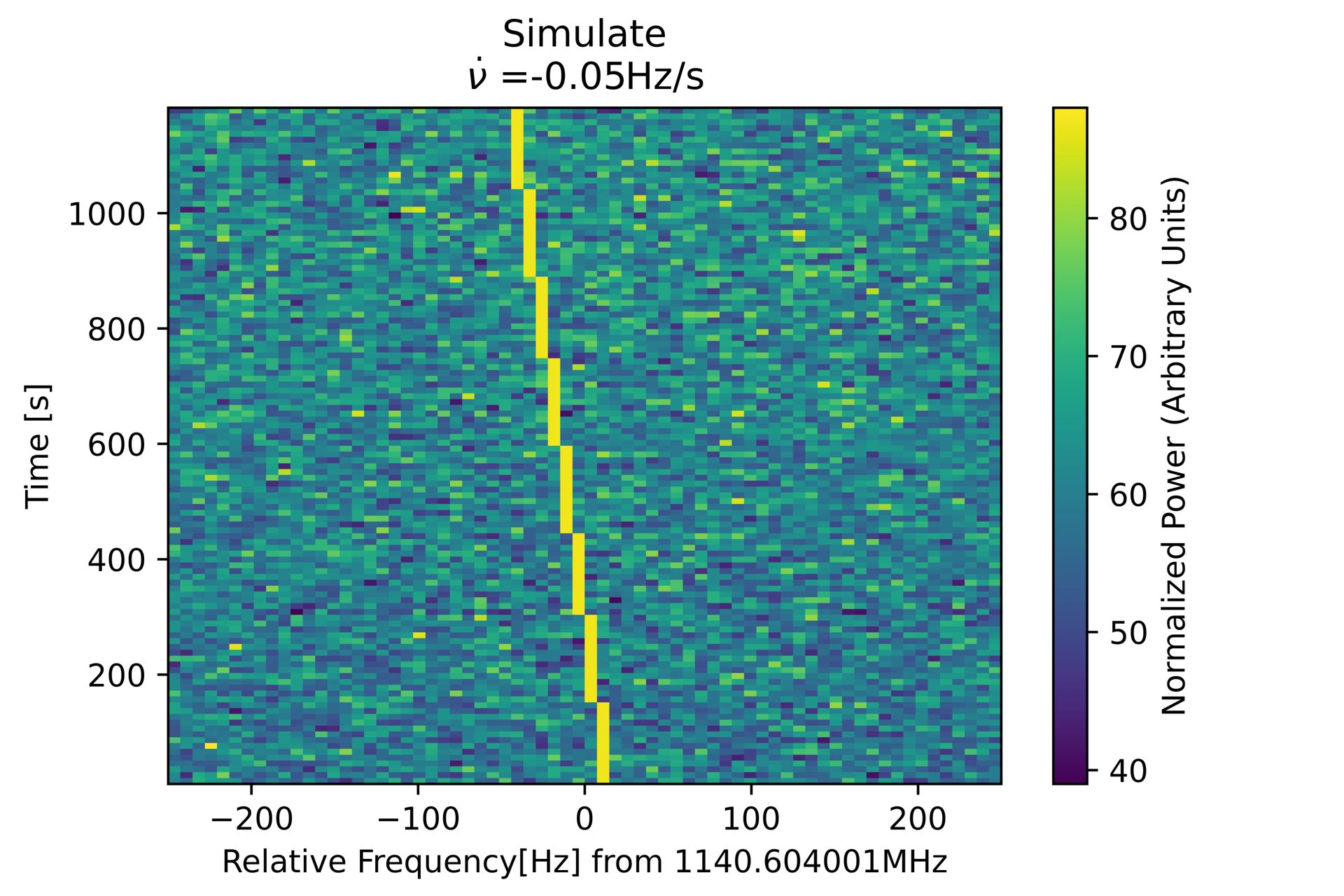}
           \caption{A simulated short-term observation for a radio source with $\dot{\nu}=-0.05$ Hz/s. }
           \label{Satellite}
\end{figure}

 For shorter observation periods (several minutes), the frequency variation, for most of time, can be approximately regarded as a linear function expressed as $\nu =\nu _0+\dot{\nu}t$, where the drift rate can be regarded as a constant (See Figure \ref{Satellite}). Since the observation periods are significantly shorter than the orbital periods of exoplanets, the possibility that the observation periods coincide with the phases of the peaks or troughs is less than other phases. Particularly, in spite of the stationary frequencies of signals from satellites, the received signals may still vary overtime, as the satellites may slowly pass through the observation area of the telescope, leaving linearly varying signals in the observational results similar to Figure \ref{Satellite}. 
\par Any drift rate smaller than the minimum drift rate $\dot{\nu}_{\min}$ is indistinguishable from the zero drift rate. $\dot{\nu}_{\min}$ can be written as (\citealt{2019ApJ...884...14S})
\begin{equation}
    \dot{\nu}_{\min}=\frac{\Delta \nu_\text{bin}}{t_\text{obs}},
\end{equation}
where $\nu_\text{bin}$ is the size of the frequency bin, $t_{\text{obs}}$ is the duration of the observation. To see tiny drift from the result, the drift rate of the signal should be greater than $\dot{\nu}_{\min}$. That is, the frequency resolution of the telescope should be grater than $\Delta \nu_\text{bin}$.
\par
\begin{figure}[htb]
  \centering
          \centering
           \includegraphics[scale=0.6]{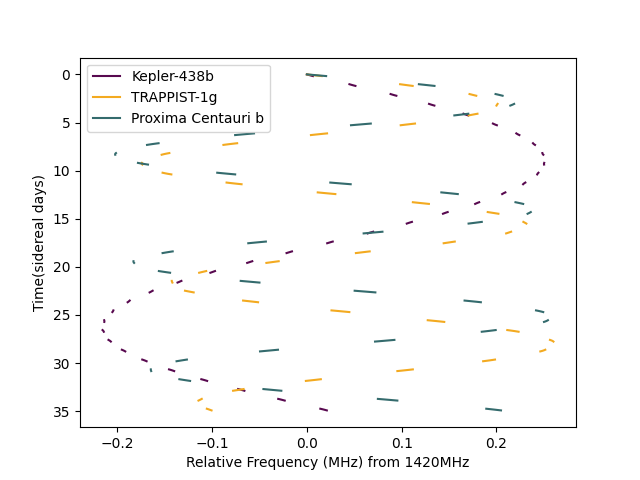}
           \caption{Intermittent pseudosinusoidal curves of simulated long-term observations for Kepler-438 b, TRAPPIST-1 g and Proxima Centauri b. Every observation lasts for approximately four hours.}
           \label{intermittent}
\end{figure}

 Theoretically, if the observation period for an exoplanet is long enough, the resulting ETI signal should completely or partly exhibit a pseudosinusoidal curve (See Figure \ref{exoplanets}). However, the signal can only be received when we are in the hemisphere facing the observation target. The final pseudosinusoidal curves are likely to be intermittent (See Figure \ref{intermittent}), unless the orbital periods of exoplanets are shorter than the observation period.
 \par
\begin{figure}[htb]
          \centering
           \includegraphics[scale=0.4]{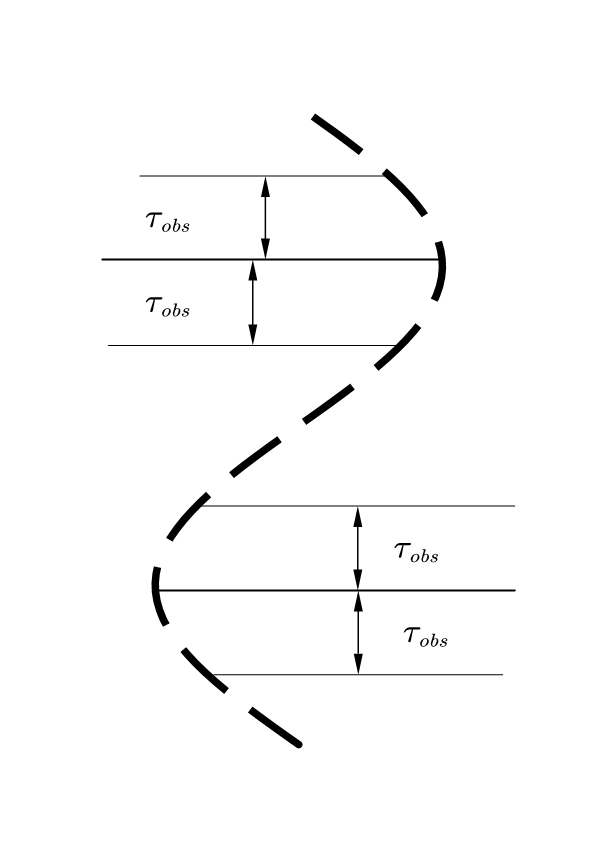}
           \caption{The situation that observation time $\tau_{\text{obs}}$ coincides with the phase of peak or trough part. For a pseudosinusoidal curve, the monotonicity changes in the peak or trough part.}
           \label{obsperiod}
\end{figure}

 What we expect to see is the peak or trough parts of the curves, since the other parts still appear linear characteristics. In such cases, if the observation period $\tau_{\text{obs}}$ is shorter than a quarter of the period of motion $P$, it should coincide with the phase of peak or trough part (See figure \ref{obsperiod}), and the corresponding probability is 
\begin{equation}
p=\frac{4\tau_{\text{obs}}}{P}.
\end{equation}
The longer the observation time is, the greater the probability will be. 
\par And in the case in which the observation period $\tau_{\text{obs}}$ is longer than a quarter of the period of motion $P$, or even longer than $P$, just as Figure \ref{intermittent} shows, we are likely to observe such intermittent signal curves, or the change of monotonicity in the peak or trough parts if the signal comes from an exoplanet.
The pseudosinusoidal curve criterion and the idea of prolonging observation period will not be applied to the first observation, since it is usually a survey for plenty of targets with a shorter observation period. If the observation period $\tau_{\text{obs}}$ is long enough, we recommend arranging an amount of observing period in the follow-up targeted observation to determine potential ETI signals or signals of interest detected in the survey. If there is no signal detected in the survey or the observation time is relatively short, it is not necessary to spend such a long time for observation. We also recommend combining other science goals to increase the efficiency of the telescope resources since the probability to detect ETI signal is very low. It can also increase the possibility of detecting other astrophysical phenomena from the observation.

\par Orbits with higher eccentricities can make ETI signals more distinguishable owing to the asymmetry of the frequency variations. The accelerations are bigger when approaching the periapsis, while they get smaller when approaching the apoapsis. Such an inequality of accelerations is reflected in the asymmetry of frequency variations. And it may also be distinguished from RFI, especially from common narrowband RFI, which appear to be symmetric over time.
 \section{Conclusion}\label{Conclusion}
 \par For the commonly studied case, the drift rate of the signal is caused by the rotational and orbital movements of Earth and the exoplanet. The drift rate caused by the rotations of Earth and exoplanet can be regarded as uniform circular motion. And the drift rate caused by the orbits of Earth and exoplanet can be regarded as elliptical motion. In the previous studies, the orbital motions were considered as uniform. However, the angular velocity, as the equations derived in Section \ref{sec:CMFDR} illustrate, is nonuniform in an elliptical orbital motion. Such nonuniform orbit motion can be explained by the Kepler equation, and the Bessel's solution of Kepler equation can be extended into Fourier sine series. Apart from the celestial mechanics factors, The latitude of telescope, the decl. and the ecliptic latitude of the target, and the inclination of the exoplanet should be also taken into consideration.
 \par Then, the equations derived are used in the calculations for the cases of potential habitable planets: TRAPPIST-1 g, Kepler-438 b and Proxima Centauri b, and FAST is the receiver on the Earth. We also discuss the case for interferometers, signals from GC, RFI, spacecraft, space probes and interstellar probes in Section \ref{Different Situations}. Finally, a discussion on the effects of observation periods, orbital periods, and the eccentricities of exoplanets is carried out in Section \ref{Discussion}. 
 \par Based on Section \ref{Different Situations} and Section \ref{Discussion} we can conclude the following:
 \begin{enumerate}
     \item ETI signals can be distinguished by the intermittent and asymmetric pseudosinusoidal curve of the frequency drift. The intermittency of the curve is the fact that the targets can be only detected when we face them, and the asymmetry of the curve is caused by the high eccentricities of the exoplanets' orbits.
     \item Longer observation time and shorter orbital periods of exoplanets can increase the possibility of seeing the intermittent curves, especially the peak or trough parts, or even the complete pseudosinusoidal curves. 
     \item This method can also be applied to the observations of interferometers in the future, and the phase difference of the signal detected by different telescopes can be a easy way to distinguish ETI signals.
     \item Long-term targeted observation can be arranged if there are potential ETI signals or signals of interest found in the SETI survey. Including other science goals can enhance the efficiency of the telescope and increase the possibility for more science outcomes.
 \end{enumerate}

 ~\\
 \par We thank the anonymous referees for the comments that helped us greatly improve this article.  And thank Dan Werthimer for his great suggestions. This work was supported by the supported by the National Key R \& D Program of China under Grant No. 2018YFA0404204 and National Science Foundation of China under Grants No. 11929301. 
\bibliography{sample631}
\bibliographystyle{aasjournal}

%% For this sample we use BibTeX plus aasjournals.bst to generate the
%% the bibliography. The sample631.bib file was populated from ADS. To
%% get the citations to show in the compiled file do the following:
%%
%% pdflatex sample631.tex
%% bibtext sample631
%% pdflatex sample631.tex
%% pdflatex sample631.tex

%% This command is needed to show the entire author+affiliation list when
%% the collaboration and author truncation commands are used.  It has to
%% go at the end of the manuscript.
%\allauthors

%% Include this line if you are using the \added, \replaced, \deleted
%% commands to see a summary list of all changes at the end of the article.
%\listofchanges

\end{document}